\documentclass[twocolumn]{emulateapj}
\usepackage{multirow}
\usepackage{graphics}
\usepackage{amsmath, amsthm, amssymb}
\usepackage{natbib}

\usepackage{graphicx}
\usepackage[dvipsnames]{xcolor}

\begin{document}

\title{Exploring the origin of moving groups and diagonal ridges by simulations of stellar orbits and birthplaces
}

\author{Douglas A. Barros$^{1}$}\email[e-mail: ]{douglas.barros@alumni.usp.br}
\author{Angeles P\'erez-Villegas$^{2}$}\email[e-mail: ]{mperez@iag.usp.br}
\author{Jacques R. D. L\'epine$^{2}$}\email[e-mail: ]{jacques.lepine@iag.usp.br}
\author{Tatiana A. Michtchenko$^{2}$}\email[e-mail: ]{tatiana.michtchenko@iag.usp.br}
\author{Ronaldo S. S. Vieira$^{3,4}$}\email[e-mail: ]{ronaldo.vieira@ufabc.edu.br}

\affiliation{$^1$Rua Sessenta e Tr\^es, 125, Rio Doce, Olinda, 53090-393 Pernambuco, Brazil\\
$^2$Universidade de S\~ao Paulo, IAG, Rua do Mat\~ao, 1226, Cidade Universit\'aria, 05508-090 S\~ao Paulo, Brazil\\
$^3$Department of Applied Mathematics, State University of Campinas, 13083-859 Campinas, SP, Brazil\\
$^4$Centro de Ci\^encias Naturais e Humanas, Universidade Federal do ABC, 09210-580 Santo Andr\'e, SP, Brazil\\
}

\date{\today}

\begin{abstract}
The present paper is the culminating one of a series aimed to contribute to the understanding of the kinematic structures of the solar neighbourhood (SN), explaining the origin of the Local Arm and relating the moving groups with the spiral-arms resonances in the disk. With a model for the Galactic potential, with the Sun inside the spiral corotation resonance (CR), we integrate the 2D orbits of test particles distributed in birthplaces along the main spiral arms, the Local Arm, and in the axisymmetric disk. A comparison of the resulting $U$--$V$ plane of the SN with that provided by \textit{Gaia} DR2 confirms our previous conclusion that the moving groups of Coma Berenices, Pleiades, and Hyades are associated with the CR, and that the Hercules stream is formed by the bulk of high-order inner Lindblad resonances. The kinematic structures result from stellar orbits trapped by the spiral resonances in a timespan of $\sim 1$\,Gyr, indicating the long-living nature of the spiral structure and challenging recent arguments in favor of short-lived structures originated from incomplete phase mixing in the Galactic disk. As a byproduct, our simulations give some insight into the birthplaces of the stars presently located in the SN; the majority of stars of the main moving groups and the Hercules stream were likely born in the Local Arm, while stars of the Sirius group possibly originated from the outer segment of the Sagittarius-Carina arm. We also propose the spiral resonances as the dynamical origin for the diagonal ridges in the Galactic distribution of rotation velocities.
\end{abstract}


\maketitle

\section{Introduction}
\label{sec:intro}

A plethora of kinematic substructures in the stellar velocity distribution of the solar neighbourhood (SN) has been unveiled with unprecedented detail by the \textit{Gaia} mission second data release (DR2, \citealt{Gaia2018A}). Consequently, these new data allowed to detect very well the richness of substructures in the stellar $U$--$V$--velocity distribution of the SN, such as the moving groups, streams, arch-like features, etc. \citep{Antoja_etal2018}. \textit{Gaia} DR2 was preceded and followed by the publication of dynamical models aiming to explain the structures on the $U$--$V$ plane of the SN. Most of the models proposed that the Sun's position coincides with some resonance of the Galactic bar \citep[e.g.][]{Dehnen2000, Minchev_etal2007, Antoja_etal2014, Monari_etal2017, Perez_Villegas_etal2017}, where, in general, they tried to explain the bimodality observed in the velocity distribution formed by the bulk of moving groups (i.e., the Coma Berenices, Pleiades, Hyades, and Sirius groups) and the Hercules stream. On the other hand, there are some models that focused mainly on the role of spiral arms in the SN velocity distribution \citep[e.g.][]{Antoja_etal2011, Quillen_etal2018, Hunt_etal2018, Michtchenko_etal2018b, Hattori_etal2018}.

The present work is the last one of a series of papers that focus on plausible mechanisms responsible for the formation of moving groups and the density structures on the $U$--$V$ plane of the SN, cf. \cite{Michtchenkoetal2017AA,Lepine_etal2017,micht,Michtchenko_etal2018b,Michtchenko_etal2019}.
 
The model of \citet[][hereafter Paper I]{Michtchenko_etal2018b} proposed to associate the different structures in the $U$--$V$ distributions with the spiral corotation resonance (CR) and the high-order Lindblad resonances (LRs), accompanying closely the CR in phase space. Its initial success in explaining the substructures prompted us to further explore that model, employing a quite different method that consisted of numerical simulations of stellar orbits under the Galactic gravitational potential and with some specific initial conditions. The present paper presents the results of this analysis, which fully supports the results of Paper I and provides further information on the initial positions and velocities of the stars that form the velocity substructures on the $U$--$V$ plane of the SN.

In addition to the very detailed velocity distribution in the SN shown by \textit{Gaia} DR2, it also revealed the presence of diagonal ridge-like structures in the distribution of Galactocentric tangential velocities as a function of Galactic radius \citep{Antoja_etal2018}. Many models and simulations have been devoted to explain the dynamical mechanism responsible for the formation of those structures. Some of them are: phase mixing in the horizontal direction \citep{Antoja_etal2018}; perturbation from the central bar \citep{Antoja_etal2018,Fragkoudi_etal2019} and from a combination bar+spiral arms \citep{Martinez-Medina_etal2019}; interaction between the Milky Way and the Sagittarius dwarf galaxy \citep{Laporte_etal2019,Khanna_etal2019}. 
In the present work, we show that the diagonal ridges could be the result of the stellar resonant orbital structure induced by the spiral perturbation, which creates regions in phase space of stable and unstable orbits. 
The moving groups and arches in the $U$--$V$ velocity space of the SN are projections of the diagonal ridges, at Galactic positions close to the solar radius \citep{Ramos_etal2018,Antoja_etal2018}.

This paper is organized as follows. In Section~\ref{sec:data_model}, we describe the data samples selected from \textit{Gaia} DR2 and present the details of the Galactic model used for the integration of the orbits. In Section~\ref{sec:simulat}, we present the recipe followed to construct the initial conditions employed in the simulations. In Section~\ref{sec:dens_struct}, we present the $U$--$V$ planes resulting from the simulations and compare them with the one obtained with \textit{Gaia} data. The diagonal ridges on the $R$--$V_{\varphi}$ plane resulting from the simulations are presented in Section~\ref{sec:ridges}, and the birthplaces of the stars that produce the current $U$--$V$ picture of the SN are discussed in Section~\ref{sec:birthplaces}. A final discussion and concluding remarks are drawn in the closing Section~\ref{sec:conclusion}.


\section{Data and Galactic Model}
\label{sec:data_model}

\textit{Gaia} DR2 provides 6D phase-space coordinates for 7,224,631 stars. Our sample is restricted to stars with relative parallax errors smaller than $20\%$ (6,376,803 stars). The heliocentric distances and the heliocentric $U$ and $V$ velocities of the stars are calculated following the formalism described in \cite{Johnson_Soderblom1987}, being $U$ positive towards the Galactic Center and $V$ positive towards the direction of Galactic rotation.

\begin{figure}
	\includegraphics[width=1.0\columnwidth]{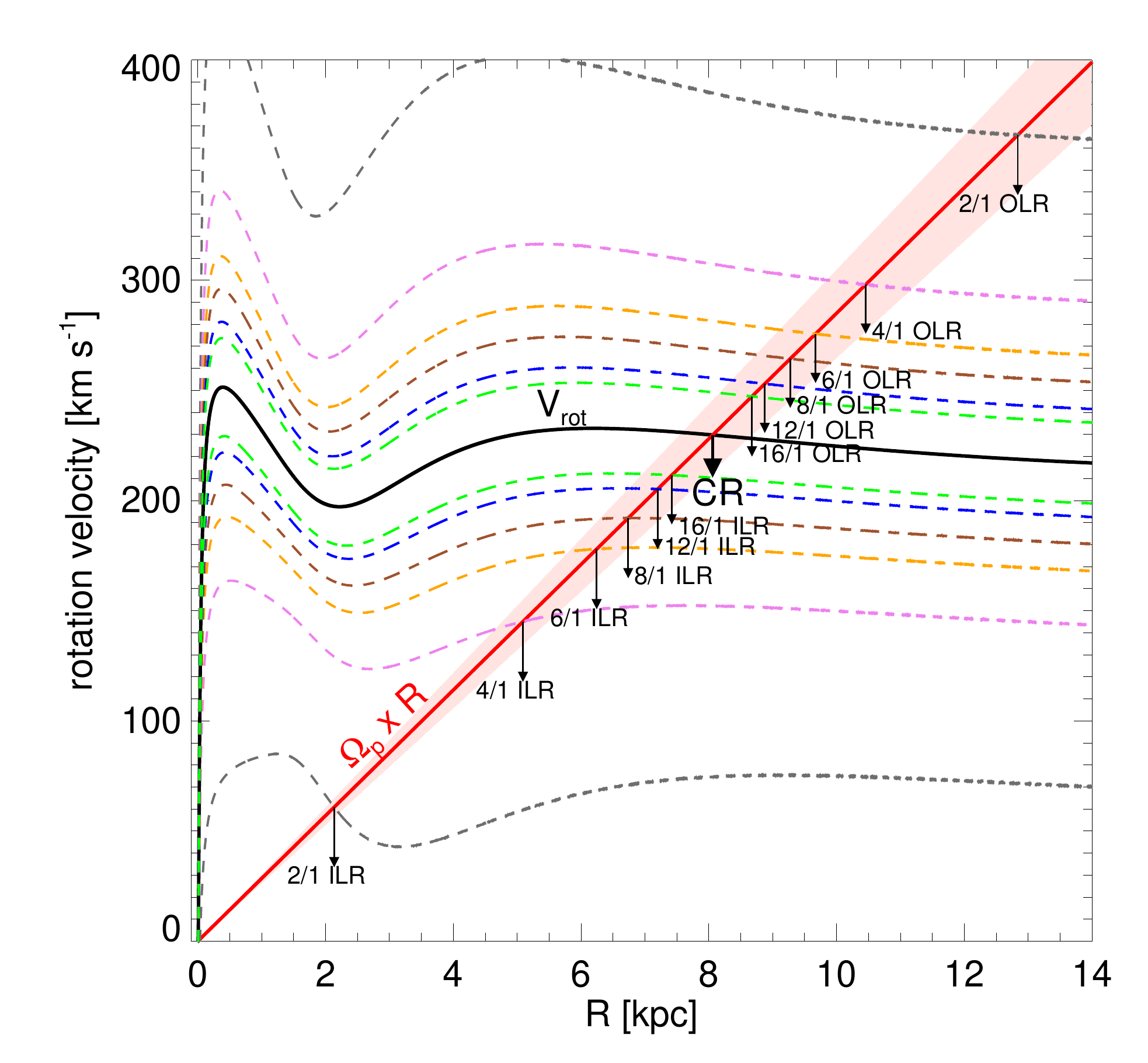}
    \caption{Rotation velocities as a function of Galactic radius from the adopted Galactic model. The black curve indicates the rotation curve, $V_{\mathrm{rot}}$, from Eq.~\ref{eq:Vrot}. The slope of the red straight line indicates the spiral-arm pattern speed, and the light shaded region relates to an uncertainty on $\Omega_{\mathrm{p}}$ of 2\,km\,s$^{-1}$\,kpc$^{-1}$, as obtained by \cite{Dias_etal2019}. The dashed curves represent the function \mbox{$(\Omega \pm \kappa/n)\times R$}; for $n= 2$ (gray), $4$ (violet), $6$ (orange), $8$ (brown), $12$ (blue), and $16$ (green).
    The arrows indicate the nominal positions of the main LRs, as well as of the CR.
    }
    \label{fig:rotcurve}
\end{figure}

The physical basis of the Galactic model applied in the present work is relatively simple \citep[see][hereafter Paper II, for more details]{Lepine_etal2017}. The axisymmetric background of the Galactic potential is made by the sum of a bulge, a disk, and a halo,
derived solely from the observed rotation curve, which is fitted by the function \citep[for details, see][]{Michtchenkoetal2017AA}:
\begin{equation}
V_{\mathrm{rot}}(R)=298.9\,\mathrm{e}^{-\left(\frac{R}{4.55}\right)-\left(\frac{0.034}{R}\right)} + 219.3\,\mathrm{e}^{-\left(\frac{R}{1314.4}\right)-\left(\frac{3.57}{R}\right)^{2}}\,,
\label{eq:Vrot}
\end{equation}
where $R$ is the Galactocentric distance given in kpc, and $V_{\mathrm{rot}}$ is the rotation velocity given in km\,s$^{-1}$. The Sun is placed at $R_{0}=8.0$ kpc, with a circular velocity at the local standard of rest (LSR) of $V_{0}=230$\,km\,s$^{-1}$. The Galactic rotation curve fitted by the function from Eq.~\ref{eq:Vrot} is represented by the solid black curve in Figure~\ref{fig:rotcurve}.

To this background, we superimposed a four-armed spiral model, which has Gaussian-shaped azimuthal profiles \citep[][]{junqueiraEtal2013AA,Michtchenkoetal2017AA}.
The spiral-arms potential is given by
\begin{equation}
 \Phi_{\rm sp}(R,\varphi) = - \zeta_0\,R\,\mathrm{e}^{-\frac{R^2}{\sigma^2}[1-\cos(m\varphi-f_m(R))]-\epsilon_s R}\,,
\label{eq:Phi_s}
\end{equation}
where $\varphi$ is the Galactocentric azimuthal angle. The parameters of the spiral arms are the same as those used in Paper II: the number of arms $m=4$, the arm width \mbox{$\sigma |\sin(i)|=1.94$}\,kpc, where \mbox{$i=-14^{\circ}$} is the pitch angle, the radial scale-length \mbox{$\epsilon_{s}^{-1}=4.0$}\,kpc, and the spiral-arm strength \mbox{$\zeta_{0}=200.0$}\,km$^{2}$\,s$^{-2}$\,kpc$^{-1}$.
The geometry of the arms is given by the function \mbox{$f_m(R)=\dfrac{m}{|\tan(i)|}\ln{(R/R_i)}+ \gamma$}, where \mbox{$R_i=8.0$}\,kpc is a reference radius and \mbox{$\gamma=237^\circ.25$} is a phase angle, whose values define the orientation of the spirals with respect to the Sun. Figure~\ref{fig:sim_XY_0} (top panel) shows a sketch of the spiral-arms model; the orange logarithmic spirals represent the loci of minima of the spiral potential from Eq.~\ref{eq:Phi_s}.

We adopt the pattern speed of spiral structure as \mbox{$\Omega_{\mathrm{p}}=28.5$}\,km\,s$^{-1}$\,kpc$^{-1}$ and the resulting corotation at the radius  \mbox{$R_{\mathrm{CR}}=8.06$}\,kpc (Paper II), which places the Sun close to the spiral corotation radius \citep{Dias_etal2019}. 
The red straight line in Figure~\ref{fig:rotcurve}, whose equation is \mbox{$\Omega_{\mathrm{p}}\times R$}, shows the rotation pattern of the spiral structure as a rigid body; the slope of the red line indicates the spiral-arm pattern speed, $\Omega_{\mathrm{p}}$. 
The dashed curves represent the function \mbox{$(\Omega \pm \kappa/n)\times R$}, with \mbox{$n=\{2,4,6,8,12,16\}$}, and $\Omega$ and $\kappa$ being the angular and epicyclic frequencies, respectively. According to the epicyclic approximation, the LRs occur when \mbox{$\Omega \pm \kappa/n=\Omega_{\mathrm{p}}$}; the inner Lindblad resonances (ILRs) and outer Lindblad resonances (OLRs) correspond to the negative and positive signs, respectively. With the adopted rotation curve and the pattern speed of the spiral arms, the main LRs are situated at the following radii, in kpc: ILRs (2/1, $R=2.14$;  4/1, $R=5.09$; 6/1, $R=6.25$; 8/1, $R=6.74$; 12/1, $R=7.20$; 16/1, $R=7.42$), OLRs (16/1, $R=8.68$; 12/1, $R=8.88$; 8/1, $R=9.28$; 6/1, $R=9.67$; 4/1, $R=10.46$; 2/1, $R=12.84$). One can see that there are eight important resonances in a distance range of 2\,kpc around CR. 

In the present work, we do not consider the perturbations from the central bar in the Galactic model. 
The reason for that is because we intend to isolate the effects of the spiral arms in the SN.
Moreover, \citet[][their Figure 13]{micht} show that, for adequate combinations of mass and pattern speed values of the bar, the corotation zone of the spiral structure is stable, and that is a requirement 
for the dynamical stability of the Local arm that is created by this resonance (Paper II). For instance, for a bar with a pattern speed close to 40\,km\,s$^{-1}$\,kpc$^{-1}$ \cite[e.g.][]{Portail_etal2017, Bovy_etal2019}, and a mass of the order of a few $10^9$ M$_{\odot}$ \citep{Lepine_Leroy2000}, the center of the corotation zone keeps stable and the motion in the SN is dominated by the spiral arm potential. In that case, we can neglect the effects of the bar on the dynamics of the SN.

\section{Simulations}
\label{sec:simulat}

We perform numerical integration of test-particle orbits in the plane of the Galaxy, using the gravitational potential model described in Section~\ref{sec:data_model}.
The total number of star-particles used is $\sim 1.4\times 10^{8}$. The total timespan of integration is 3\,Gyr, which corresponds to approximately 14 revolutions of the spiral pattern in the inertial frame. The integration of the equations of motion is performed by means of a fifth-order Runge-Kutta integration scheme, with a time-step of 1\,Myr. The role of the test particle integration is to allow the imposed initial conditions (positions and velocities) of the particles to adapt to the modelled gravitational force field through their orbital evolution.

In the simulations, we do not model an adiabatic introduction of the spiral-arms perturbation, but instead we fix the spiral amplitude at its current value. Models with a growing amplitude of the spiral potential are present in the literature, but they are not applicable in the present work, since we simulate an initial disk with a set of stars already distributed along the spiral arms, thus meaning a spiral potential present at the instant $t=0$. We show in the next sections that the dynamical equilibrium of the system is reached in a time-scale of order of 1\,Gyr. 


\subsection{Initial conditions}
\label{sec:init_cond}

The star-particles are initially distributed in the Galactic disk according to the following conditions:

\begin{itemize}
\item \textbf{The four main spiral arms (MSAs)} -- the radii $R$ are randomly distributed on the arms between Galactic radii 5\,kpc and 12\,kpc. The azimuths $\varphi$ follow random Gaussian distributions with mean values along the loci of minimum of the spiral potential in Eq.\,(\ref{eq:Phi_s}), while the width of the distributions is given by the $\sigma$ parameter. With these azimuths and radii, we simulate patches of the four MSAs: Sagittarius-Carina, Perseus, Scutum-Crux, and Norma-Cygnus arms.
\item \textbf{The Local Arm (LA)} -- the stars are placed inside the local corotation zone between the Sagittarius-Carina and Perseus arms (see Paper II). The radii are randomly distributed along the corotation radius at 8.06\,kpc with the width of 0.4\,kpc. The azimuths are distributed in the range \mbox{$40^{\circ}\leq \varphi \leq 120^{\circ}$}.
\item \textbf{The stellar disk background} -- the radii and azimuths are chosen randomly through uniform distributions in the ranges \mbox{$5 \leq R \leq 12$}\,kpc and \mbox{$0^{\circ}\leq \varphi \leq 360^{\circ}$}, respectively. This component  simulates an initially axisymmetric disk.
\end{itemize}

\begin{figure}
	\includegraphics[width=1.0\columnwidth]{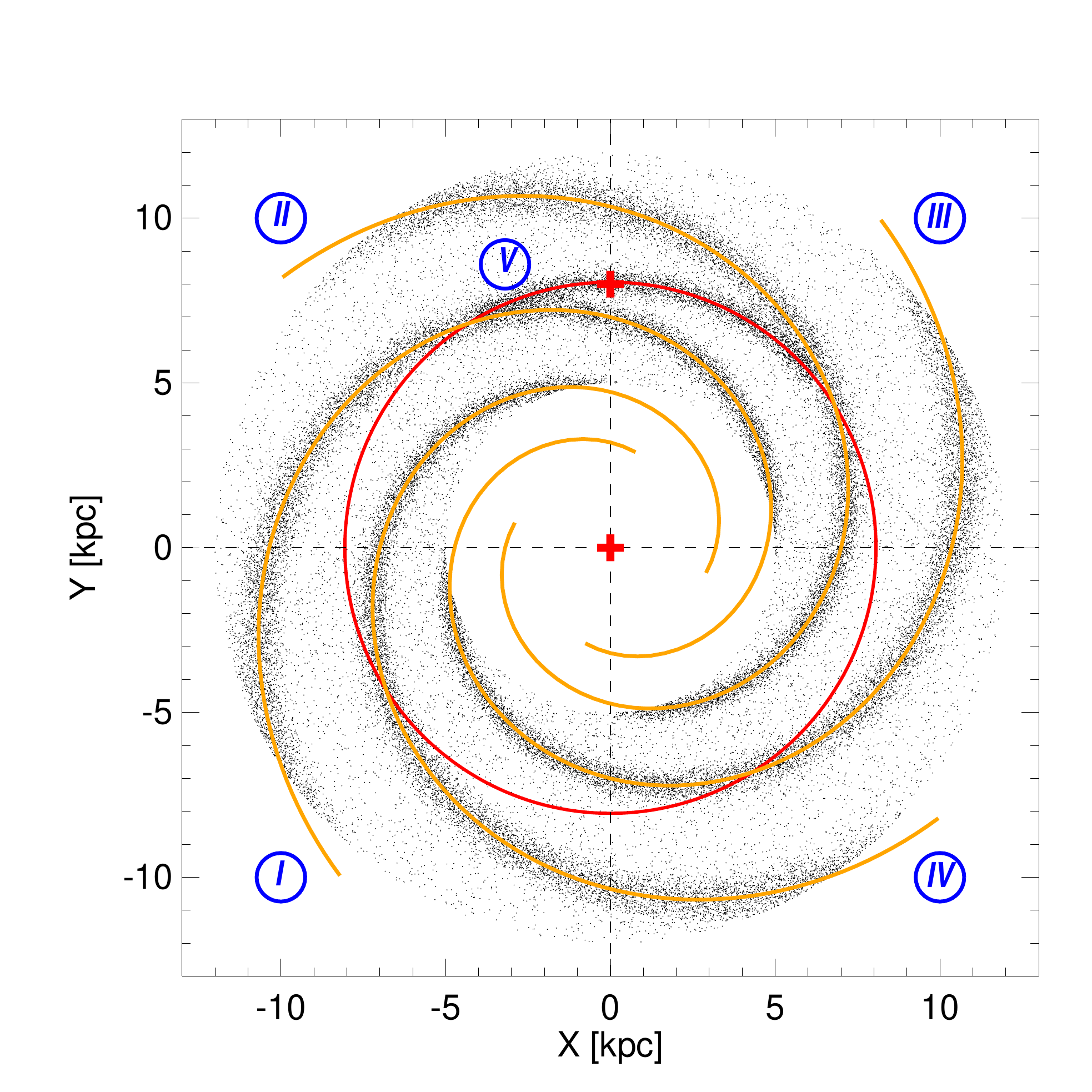}
	\includegraphics[width=1.0\columnwidth]{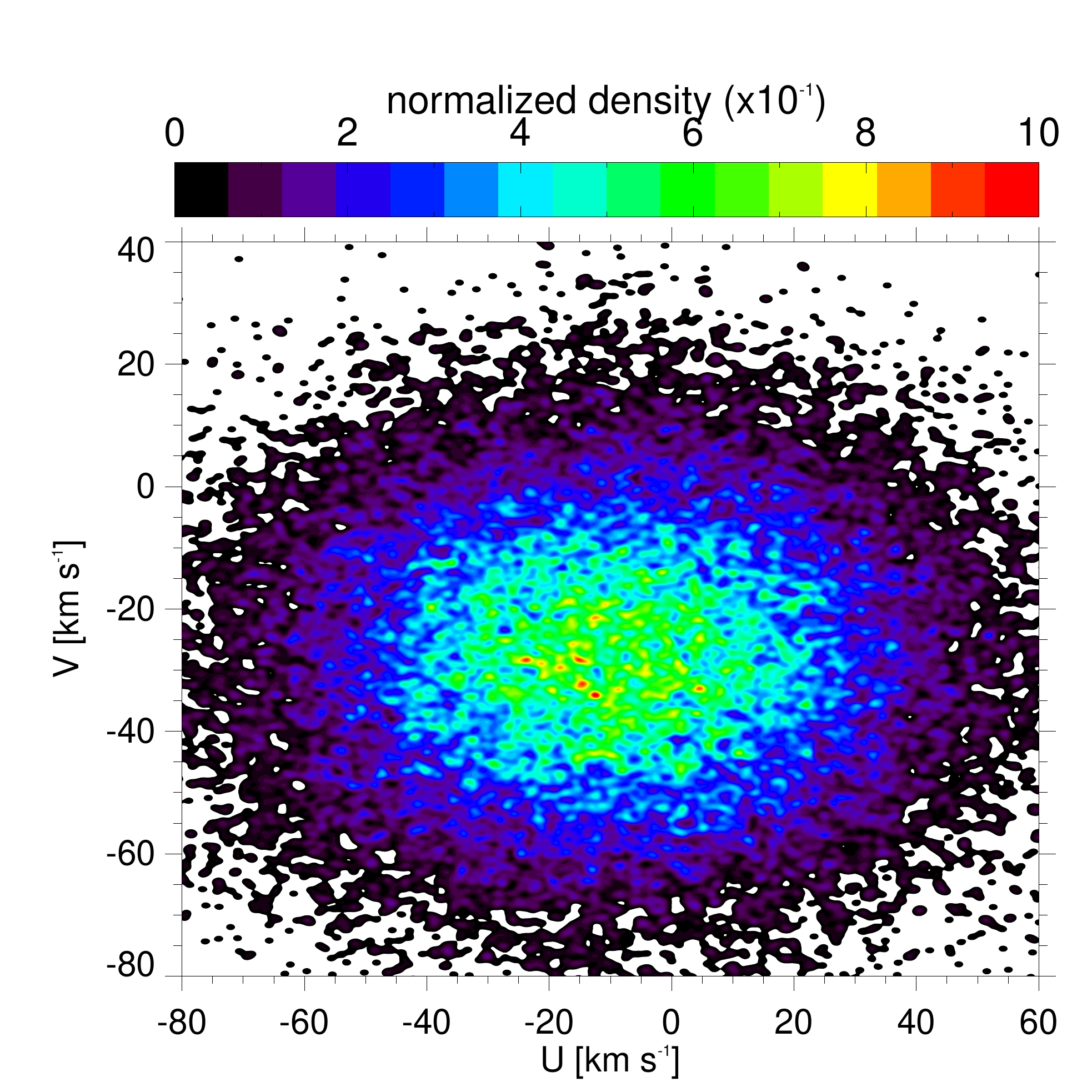}
    \caption{Top: Initial positions of the star-particles (black dots) on the Galactic equatorial $X$--$Y$ plane (about 60\,000 randomly chosen particles are plotted). The orange logarithmic spirals represent the four Galactic arms: (\textrm{I}) Sagittarius-Carina; (\textrm{II}) Perseus; (\textrm{III}) Norma-Cygnus; and  (\textrm{IV}) Scutum-Crux. The red circle represents the corotation circle, while the Local Arm (\textrm{V}) is represented by the red circular segment between the Sagittarius-Carina and Perseus arms. The red crosses indicate the positions of the Galactic Center and the Sun, at $(X,Y)=(0,0)$\,kpc and $(0,8.0)$\,kpc, respectively.
    Bottom: Distribution on the heliocentric $U$-$V$--plane, at $t=0$. The smoothed distribution of stars in the plane is obtained by a kernel density estimation technique in two dimensions, giving the normalized densities. The color scale, from violet to red, indicates the range of densities. The red peaks originate from the random velocity distributions, showing the noise level.
    }
    \label{fig:sim_XY_0}
\end{figure}

Figure~\ref{fig:sim_XY_0} (top panel) shows the initial distribution of star-particles on the Galactic equatorial $X$--$Y$ plane. The number of star-particles is distributed between the four MSAs and the LA in such a way that the stellar densities of these arms are comparable (e.g. \citealt{Xu_etal2016}). In this way, to each MSA, we attribute a number of $2.56\times 10^{7}$ test particles; to the LA, $1.024\times 10^{7}$ test particles are attributed; and finally, $2.56\times 10^{7}$ test particles are uniformly distributed in the stellar disk background. It is worth noting that the particle's number initially attributed to each one of the above components affects only the relative densities (contrasts) of resulting structures, but preserves the qualitative structure of the simulated $U$--$V$ plane.

The initial distributions of the star-particles Galactocentric velocities, radial $V_R$ and tangential $V_{\varphi}$ (with respect to the inertial frame) components, are constructed in the following way: following  \cite{Dehnen1999}, we consider the radial velocity dispersion \mbox{$\sigma_{V_{R}}^{3}\propto \Sigma$}, where $\Sigma(R)$ is the disk surface density. For a disk with an exponential radial density profile, it gives us \mbox{$\sigma_{V_{R}}=\sigma_{0}\,\mathrm{e}^{-\frac{(R-R_{0})}{3 R_{\Sigma}}}$}. We adopt the local velocity dispersion \mbox{$\sigma_{0}=30$}\,km\,s$^{-1}$ and the scale-length \mbox{$R_{\Sigma}=2.5$}\,kpc \citep{Freudenreich1998}. The initial radial velocities $V_R$ of the star-particles have a Gaussian distribution with zero mean value and width $\sigma_{V_{R}}$ at each radius.

The profile of the tangential velocity dispersion is derived from the epicyclic approximation \mbox{$\sigma_{V_{\varphi}}^{2}/\sigma_{V_{R}}^{2}=\kappa^{2}/(4\Omega^{2})$} 
\citep[][Eq.\,4.317]{Binney_Tremaine2008}. The resulting local dispersion is \mbox{$\sigma_{V_{\varphi}}\approx 20$}\,km\,s$^{-1}$.
We also consider the asymmetric drift term $v_{a}$ \citep[][Eq.\,4.228, see also \citealt{Antoja_etal2011}]{Binney_Tremaine2008}. 
The initial tangential velocities $V_{\varphi}$ of the star-particles are then randomly sorted using a Gaussian distribution with mean value \mbox{$(V_{\mathrm{rot}}-v_{a})$} and width $\sigma_{V_{\varphi}}$. Finally, the initial values of the linear and angular momenta (per unit mass) are fixed as \mbox{$p_{R}=V_{R}$} and \mbox{$L_{z}=R\,V_{\varphi}$}, respectively.

During the integration of the orbits, we record the positions and velocities of the star-particles at each 10\,Myr. In the construction of the $U$--$V$ planes in the SN, we select star-particles within a circle of 150\,pc around the Sun located at \mbox{$(R;\, \varphi)=(8.0\,\mathrm{kpc};\, 90^{\circ})$}. The $U$ and $V$ velocities of the star-particles are calculated from the $V_R$ and \mbox{$(V_{\varphi}-V_{\mathrm{rot}})$} components and then transformed to heliocentric ones by subtracting \mbox{$U_{\odot}=11.1$}\,km\,s$^{-1}$ and \mbox{$V_{\odot}=12.24$}\,km\,s$^{-1}$ \citep{Schoenrich_etal2010}. 
For a better match of the structures on the simulated \mbox{$U$--$V$} plane with those from \textit{Gaia} DR2, we further subtract 6.3\,km\,s$^{-1}$ from the $V$-velocities. This correction corresponds to a solar velocity component of \mbox{$V\approx 18.5$}\,km\,s$^{-1}$ in the direction of Galactic rotation\footnote{We are currently analysing the LSR correction based on the \textit{Gaia} data and we have strong evidence that \mbox{$V\approx 18.5$}\,km\,s$^{-1}$ reflects the right value of the $V$-component of the solar motion.}. 
Figure~\ref{fig:sim_XY_0} (bottom panel) shows the simulated \mbox{$U$--$V$} plane at \mbox{$t=0$}; the red peaks show the noise level originated from the random velocity distributions.

\subsection{Justification for the adopted initial conditions}
\label{sec:just_init_cond}

The choice of the initial positions of the star-particles shown in Figure~\ref{fig:sim_XY_0} (top panel) is based mainly on the assumption that the stars are born predominantly in the spiral arms because of the high density in gas and dust, which is supported by observations of the Milky Way and of other galaxies. The locations of the four MSAs adopted in this work are confirmed by observations of Galactic star-forming regions \citep[e.g.][]{Hou_Han2014,Reid_etal2014} and tangent-line data \citep{Vallee2016}.

The LA is an observable structure, located midway between the Sagittarius-Carina and Perseus arms. The Sun is currently close to LA \citep{Reid_etal2014,Hou_Han2014,Xu_etal2016} and we expect that a significant part of the SN must be composed of stars coming from this structure. Paper II showed that the LA is formed by stars whose orbits are trapped inside the local corotation zone, never crossing the MSAs, but instead librating inside the region between them. This implies that the kinematics of the SN must be significantly affected by the resonant dynamics of the local corotation zone.

The choice of the initial velocities is based on results of recent studies, which show the broad velocity ranges of the newly formed objects. \citet{Wu_etal2014} and \citet{Sakai_etal2015} show that, in star-forming regions in the Sagittarius and Perseus arms, peculiar velocities vary up to \mbox{$\sim 20$}\,km\,s$^{-1}$ or more. \citet[][their Figure 5]{Reid_etal2014} show large peculiar motions in the high-mass star-forming regions in the MSAs, contrary to the LA, whose sources do not present significant deviations from circular motion. \citet[][their Figure 11]{Lepine_etal2008} show that some recently formed open clusters present deviations from circular velocities of up to \mbox{$\sim 40$}\,km\,s$^{-1}$ or more. 

\begin{figure*}
	\includegraphics[width=1.0\columnwidth]{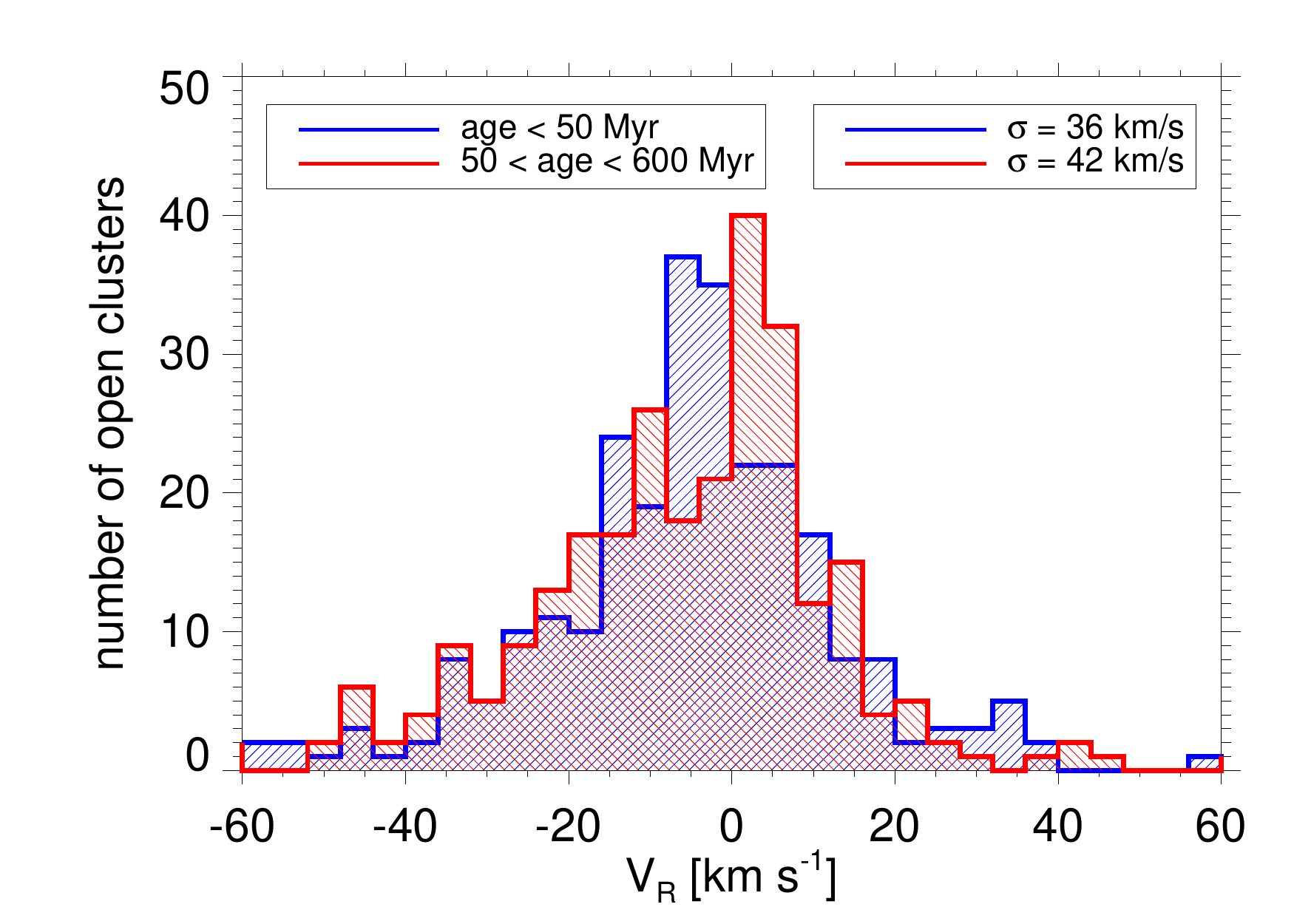}
	\includegraphics[width=1.0\columnwidth]{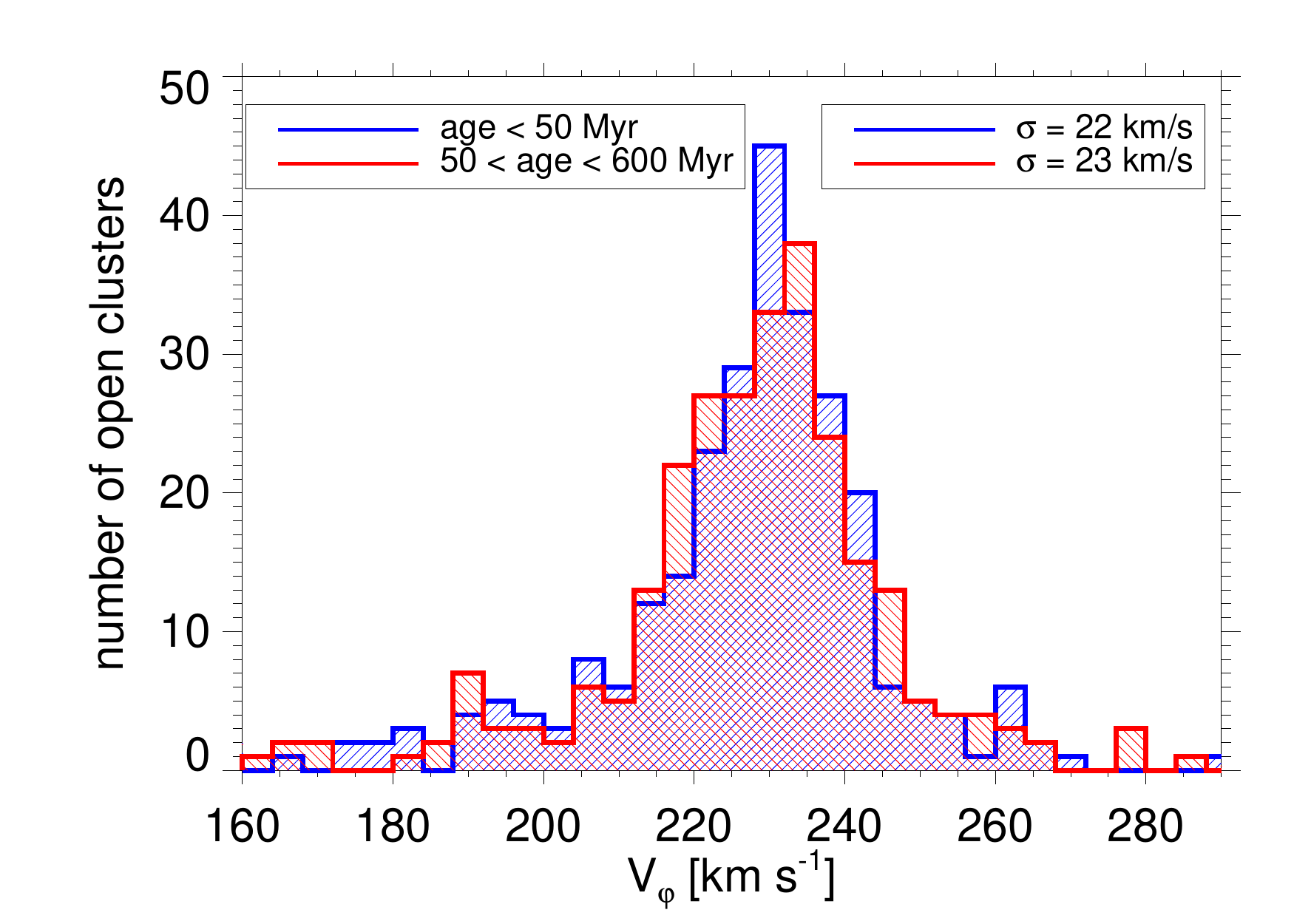}
    \caption{Histogram distributions for the Galactocentric velocities of OCs: blue for the sample with OCs younger than 50\,Myr, and red for the OCs sample with ages between 50\,Myr and 600\,Myr. Left: for the distributions of radial velocities $V_R$. Right: for the distributions of tangential velocities $V_{\varphi}$. The values of the velocity dispersion of each distribution are indicated in the legends. The bin size of all histograms is 4\,km\,s$^{-1}$.}
    \label{fig:OCs_velocities}
\end{figure*}

To get a little deeper into this issue, we analysed the velocity distributions of a sample of open clusters (OCs) from the DAML02 catalogue \citep{Dias_etal2002}. In order to do that, we selected the OCs with data on distances, proper motions, and radial velocities. In total, we collected  683 objects with the entire information. From that sample of OCs, we computed the Galactocentric radial $V_R$ and tangential $V_{\varphi}$ velocities. We took two groups with different intervals of ages and similar number of objects: one group composed by young OCs with ages $< 50$\,Myr (272 objects), which are still connected to the spiral arms where they probably were born \citep{Dias_etal2019}, and a second group with ages in the interval 50\,Myr $<$ ages $<$ 600\,Myr (274 objects), representing OCs at more advanced orbital stages. In Figure~\ref{fig:OCs_velocities}, we show the distributions of radial ($V_R$, left panel) and tangential ($V_{\varphi}$, right panel) velocities, for the two subsets of OCs. The velocity dispersion of the distributions of the OCs younger than 50\,Myr are 36\,km\,s$^{-1}$ in $V_R$ and 22\,km\,s$^{-1}$ in $V_{\varphi}$. As it can be seen from Figure~\ref{fig:OCs_velocities}, these values are similar to those presented by the velocity distributions of the OCs of intermediate ages. Also, these values are in correspondence with our adopted values for the local velocity dispersion in the initial conditions of the simulations. 

Even though the mechanisms that can produce such large deviations from circular motion of recently born stars still remain unexplained, we propose to investigate a wide range of initial velocities. In this way, we provide the conditions for stars with initial radii far from the solar radius, both inside and outside the corotation circle, to visit the SN. In fact, as shown in the next Section, the Hercules and Sirius streams in the $U$--$V$ plane are likely formed by stars that came from inner and outer regions of the Galaxy, respectively (see also Paper I).


\section{Structures on the $U$--$V$ planes}
\label{sec:dens_struct}

Figure~\ref{fig:Gaia_DR2_UV_2} shows the $U$--$V$ velocity distribution in the SN.  For the purpose of comparison, the left panel shows the velocity distribution of the \textit{Gaia} stars within 150\,pc from the Sun (195,489 stars). The right panel shows the distribution resulting from the simulation, which is constructed with $\sim$ 92,000 star-particles that fall within a circle of 150\,pc from the Sun, as described in Section~\ref{sec:init_cond}. The simulated plane is a snapshot at the instant \mbox{$t=1340$}\,Myr, which was chosen as described below. It is worth emphasizing that, in our simulations, no new generation nor death of stars are considered; all the stars are born at the beginning and remain during the entire simulation.

\subsection{Stabilization}
\label{sec:stab}

Analyzing the time evolution of the density structures on the simulated $U$--$V$ plane, at each 10\,Myr, we observe initially the stage of phase mixing in response to the spiral-arms force field \citep[e.g.][]{Antoja_etal2011}, when the structures change their sizes, shapes and positions on the plane. The process is stabilized after about \mbox{$\sim 1000$}\,Myr (roughly, one complete orbital period in the SN, in the rotating frame). Next, we start to observe only the variation of the density of the main kinematic groups, with a periodicity of \mbox{$\sim 150$}\,Myr. The snapshot of the \mbox{$U$--$V$} plane taken at \mbox{$t=1340$}\,Myr shows satisfactory similarity with the velocity plane formed by \textit{Gaia} objects. Nevertheless, for example, the planes of instants 1190\,Myr and 1490\,Myr also show a similar velocity distribution, as presented in Figure~\ref{fig:UV_planes_others}. This periodicity matches the radial variation of the stellar orbits in the SN and can be explained by the fact that no star's generation has been considered during the simulation. This result points to a picture in which the SN is composed in its majority by stars with orbital phases similar to the ones shown by the star-particles in the panels of Figure~\ref{fig:UV_planes_others}. The simulated $U$--$V$ planes after \mbox{$\sim 2000$}\,Myr and up to 3000\,Myr still present the main density structures observed on the planes of Figure~\ref{fig:UV_planes_others}, but with some lower degree of separation between the structures. It is worth emphasizing that the maximal densities (red peaks) on the simulated \mbox{$U$--$V$} plane at 1340\,Myr result from the initial random velocity distributions, in contrary to the density peaks on the \mbox{$U$--$V$} plane from \textit{Gaia} that show real stellar groupings in velocity space.

Comparing the initial velocity distribution (Figure~\ref{fig:sim_XY_0}, bottom panel) to that after roughly 1\,Gyr of the stars orbital evolution (Figure~\ref{fig:Gaia_DR2_UV_2}, right panel), we realize that the action of the spiral gravitational perturbation transforms the $U$--$V$ distribution thoroughly. The main structures that appear on the simulated $U$--$V$ plane are arches, streams, edges, etc., which have been already observed by \cite{Antoja_etal2018} in the current distribution of the \textit{Gaia} DR2 (Figure~\ref{fig:Gaia_DR2_UV_2}, left panel). Our spiral disk model provides a precise explanation for these structures, as described in Paper I. Shortly, the arch-like contours appear on the $U$--$V$ plane due to the symmetry of the Hamiltonian with respect to the momentum $p_R$ (or, correspondingly, to the velocity component $U$), the edge lines (see Section~\ref{sec:comparison}) are produced by the separatrices and saddle points of the corotation and other resonances, while streams are alternations in density between high-order Lindblad resonances and near-resonance domains.


\subsection{Comparison with observational data}
\label{sec:comparison}

\begin{figure*}
	\includegraphics[width=1.0\columnwidth]{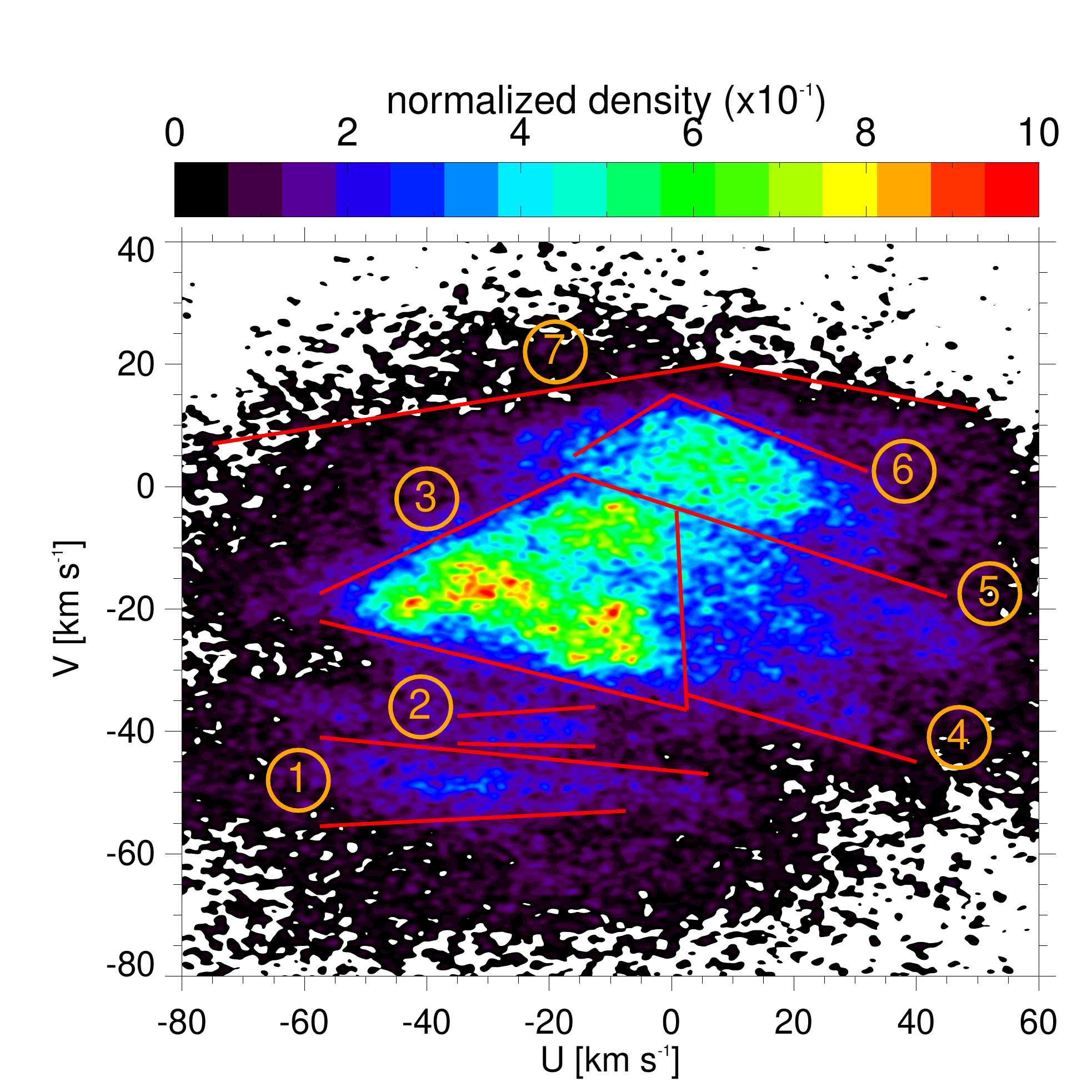}
	\includegraphics[width=1.0\columnwidth]{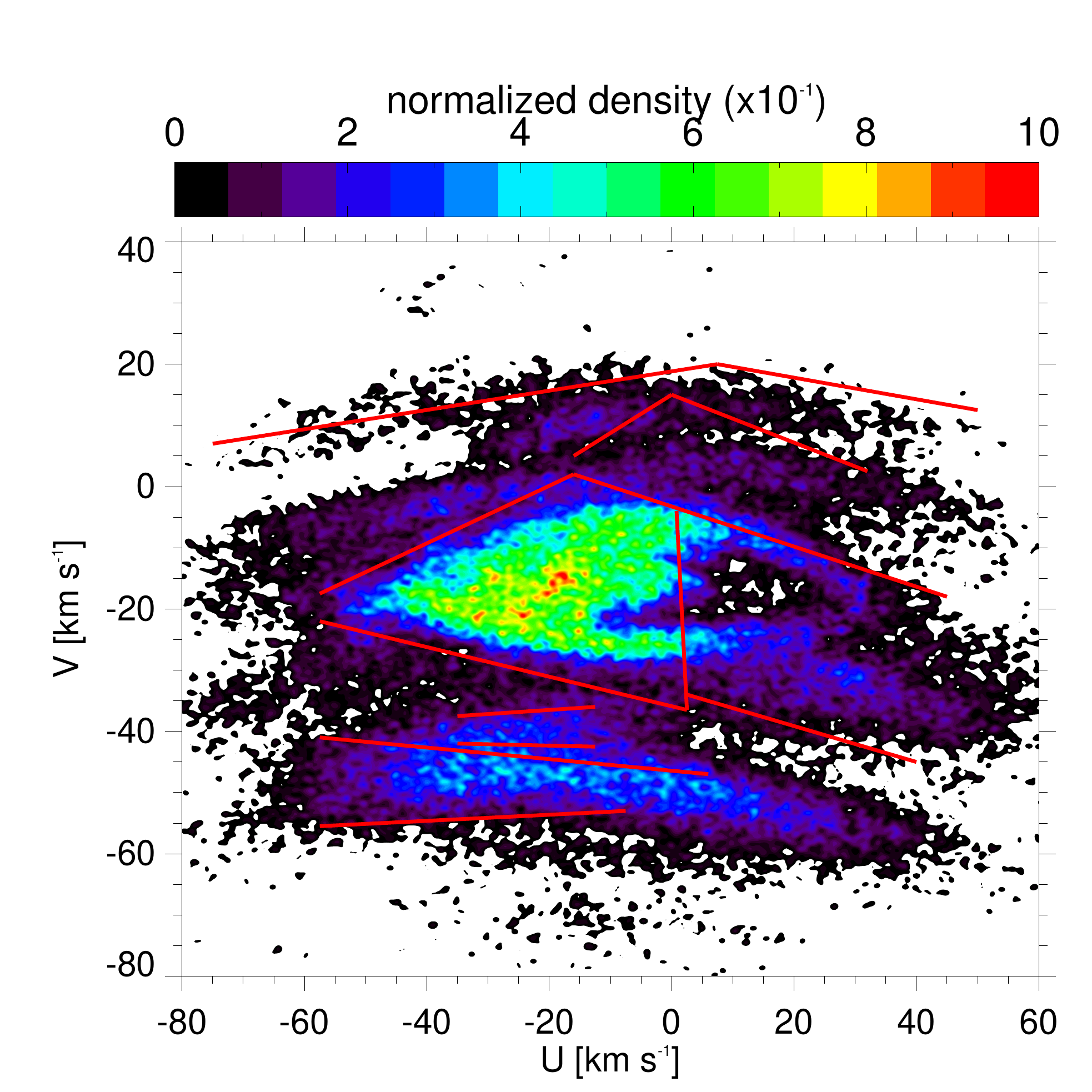}
    \caption{Left: $U$--$V$ plane of the \textit{Gaia} stars within 150\,pc from the Sun. The straight lines are ``edge lines'' (see Section~\ref{sec:comparison}). Right: $U$--$V$ plane of the SN resulting from the simulation, at $t=1340$\,Myr. The color scale indicating the normalized densities is the same as in Figure~\ref{fig:sim_XY_0}. The same set of edge lines on the left panel is overlaid.}
    \label{fig:Gaia_DR2_UV_2}
\end{figure*}

\begin{figure*}
	\includegraphics[width=0.32\textwidth]{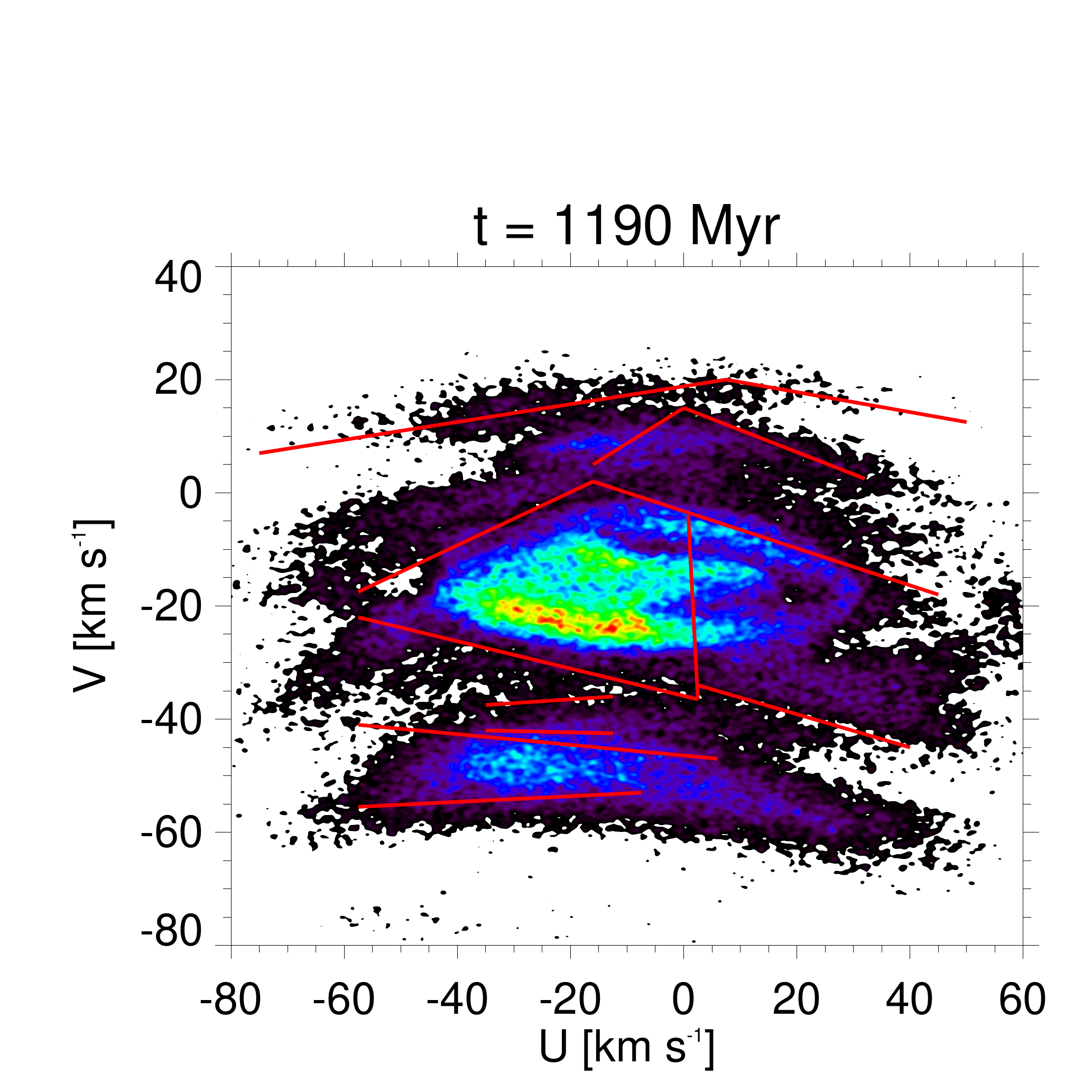}
	\includegraphics[width=0.32\textwidth]{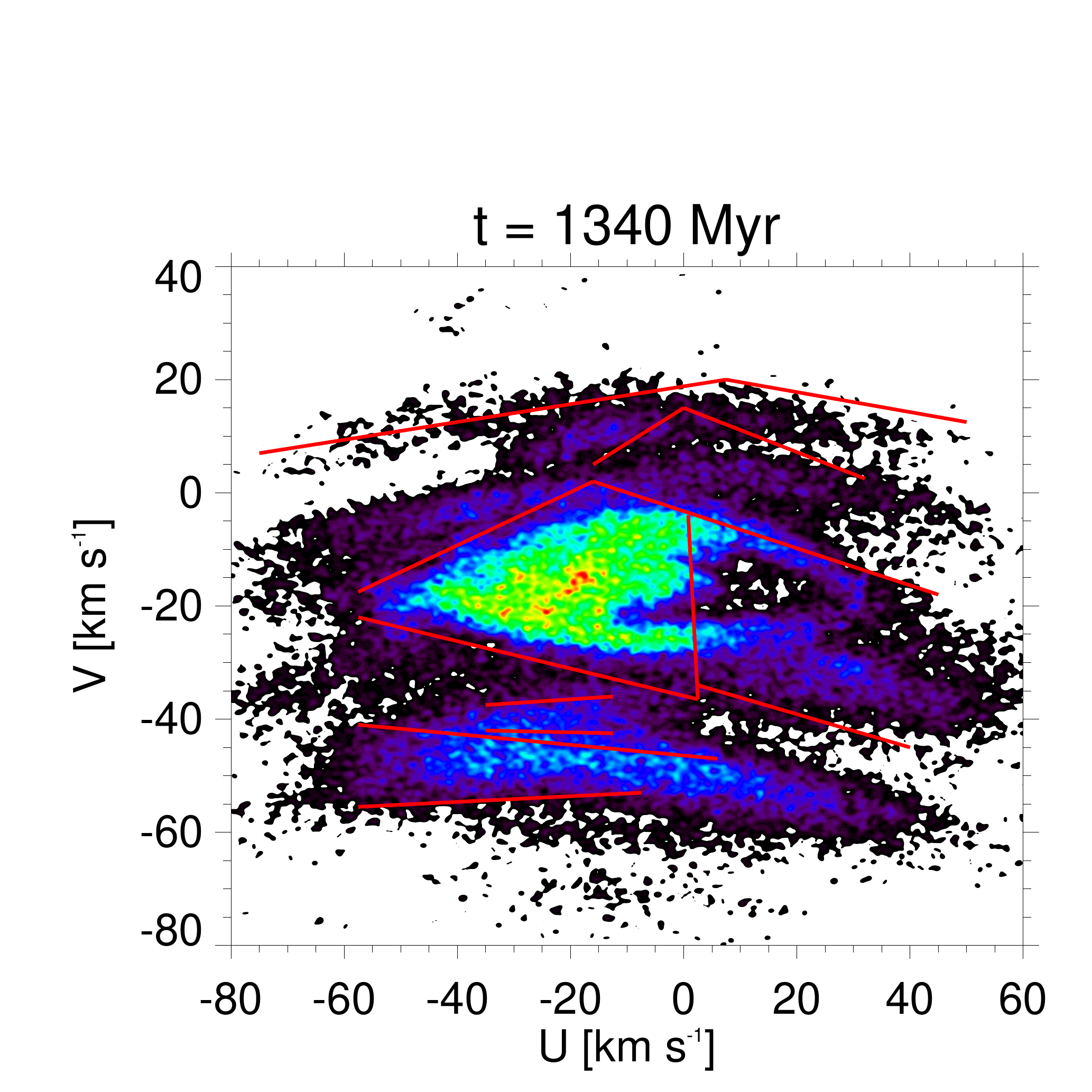}
	\includegraphics[width=0.32\textwidth]{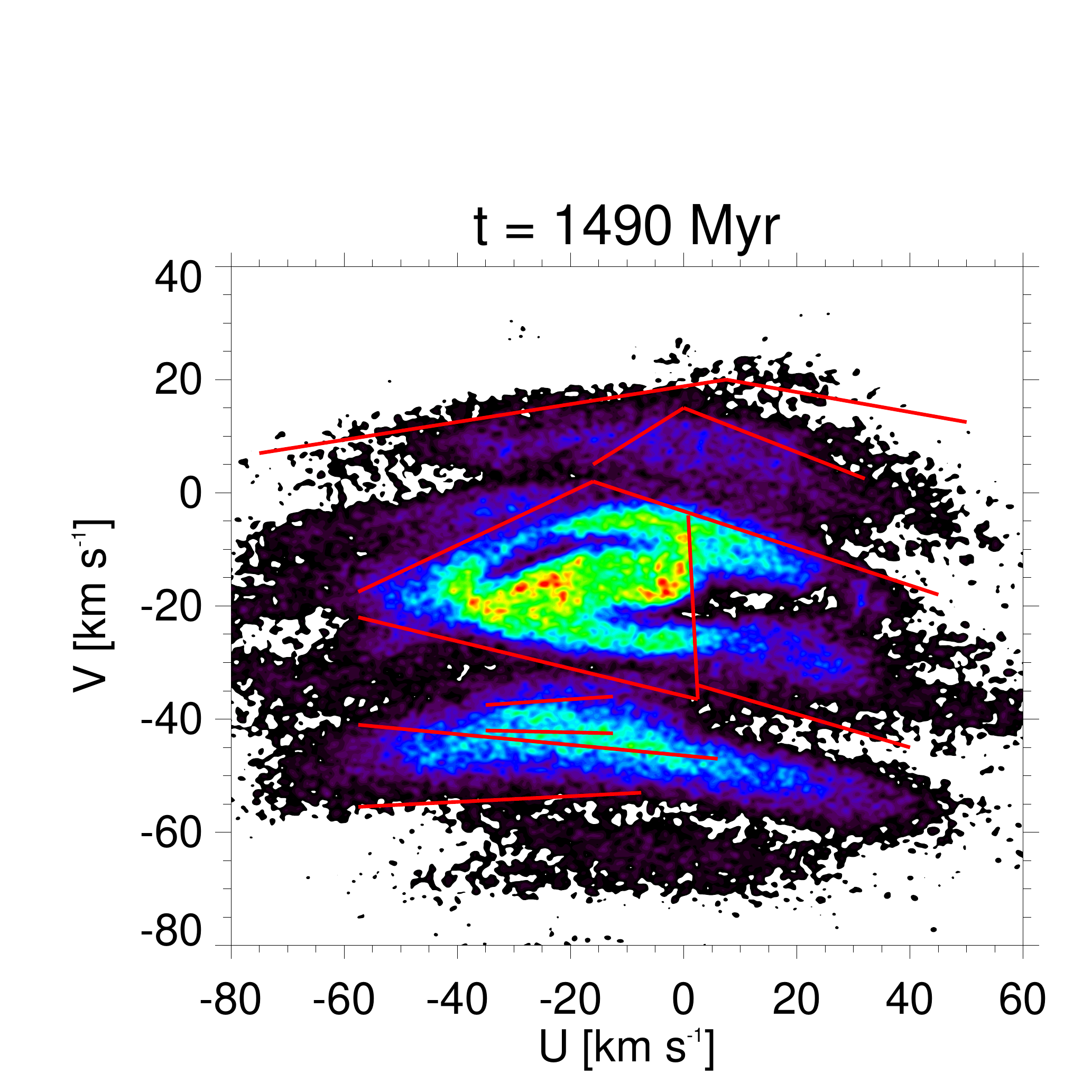}
    \caption{$U$--$V$ planes of the SN resulting from the simulation. Left: for the snapshot taken at $t=1190$\,Myr. Middle: for the snapshot taken at $t=1340$\,Myr. Right: for the snapshot taken at $t=1490$\,Myr. The color scale is the same as in Figure~\ref{fig:sim_XY_0}. The same set of edge lines on panels of Figure~\ref{fig:Gaia_DR2_UV_2} is overlaid.}
    \label{fig:UV_planes_others}
\end{figure*}

For a detailed comparison between the observed and simulated velocity distributions, we plot some ``edge lines'', identical for both $U$--$V$ planes in Figure~\ref{fig:Gaia_DR2_UV_2}. The edge lines are defined by visual inspection of the density structures observed on the \textit{Gaia} $U$--$V$ plane; we associate to each set of lines a number, from 1 to 7.  In the following, we describe the correspondence between the edge lines and the density structures in the simulated $U$--$V$ plane.

\begin{description}
\item[Edge lines \#1] we observe here a structure with density of stars similar to that shown by the data from \textit{Gaia}. This structure is associated with the Hercules stream and, according to Paper I, is formed essentially by stars that evolve inside of the 8/1 ILR.
\item[Edge lines \#2] we observe a weak density structure, which is clearly associated with the similar structure observed with the data from \textit{Gaia}. According to Paper I, this structure is formed by the stars that evolve inside of the 12/1 and higher-order ILRs, located close to the corotation zone.
\item[Quadrilateral of Edge lines \#3] the corotation zone can be observed as a density of stars similar to the one shown by the $U$--$V$ plane from \textit{Gaia}, which encompasses the moving groups of Coma Berenices, Pleiades and Hyades (the main component of moving groups). The frontiers of the abrupt variations in the stellar density (corresponding to the quadrilateral) are relatively similar to those present in the observed $U$--$V$ plane. This component is associated with the CR, whose projection on the $U$--$V$ plane appears as the region of greatest extension in $V$ velocities, as it can be seen in Paper I (their Figure 1).
\item[Edge lines \#4 and \#5] such as in the $U$--$V$ plane from \textit{Gaia}, these lines delimit a region of reduced densities in the simulated $U$--$V$ plane, but which is still associated with the CR according to Paper I (their Figure 1). The separatrices of the local corotation zone on the dynamical map of Paper I (their Figure 1), the ``Quadrilateral of Edge lines \#3'', and the ``Edge lines \#4 and \#5'' in the simulated $U$--$V$ plane, confirm our previous hypothesis that the main component of the observed $U$--$V$ plane is likely created by the zone of dynamical influence of the CR, the same that creates the spatial structure of stars that we identify with the Local Arm (Paper II). Some of the diagonal branches observed between the Edge lines \#4 and \#5 in the simulated $U$--$V$ plane are similar to the corresponding structures in the $U$--$V$ plane from \textit{Gaia}.
\item[Edge lines \#6] our simulations do not reproduce the high-density region corresponding to the Sirius moving group in the $U$--$V$ plane from \textit{Gaia}. However, we can still observe a branch of low density along \mbox{$V\simeq 5$}\,km\,s$^{-1}$ and at positive $U$ velocities, and which is likely associated with the 12/1 and 16/1 OLRs according to Paper I (their Figures 1 and 4).

\item[Edge lines \#7] arch-like upper limit of the velocity distribution in the SN. It is located up to \mbox{$V\sim 20$}\,km\,s$^{-1}$, and is associated with the 8/1 OLR (see dynamical map in Figure 1 of Paper I).
\end{description}

The region of low stellar densities with an arched shape that extends from \mbox{$(U,V)\simeq(-80,-25)$}\,km\,s$^{-1}$ to \mbox{$(U,V)\simeq(60,-60)$}\,km\,s$^{-1}$ \citep[e.g.][]{GaiaCollaboration2018}, which splits the observed $U$--$V$ plane in two major components, i.e., the Hercules stream and the main component of moving groups, has been attributed to the influence of the 2/1 OLR of the Galactic bar. A fast-rotating bar, with its 2/1 OLR close to the Sun, could generate a bimodality in the velocity plane of the SN \citep[e.g.][among others]{Dehnen2000,Minchev_etal2007,Antoja_etal2014,Monari_etal2017}. Our simulations with a long-living spiral structure, with its CR close to the Sun, also turn evident regions of low density of stars separating the main region of the Hercules stream from the main component of the moving groups. Such regions of low stellar density are localized just below the bottom line of the ``Quadrilateral of Edge lines \#3'' and below the ``Edge line \#4'' on the simulated $U$--$V$ plane. The direction in which such separation occurs, on the simulated $U$--$V$ plane, is compatible with the direction that characterizes the ``bimodality'' on the observed $U$--$V$ plane.


\section{Diagonal ridges on the $R$--$V_{\varphi}$ plane}
\label{sec:ridges}

\begin{figure*}
	\includegraphics[width=1.0\columnwidth]{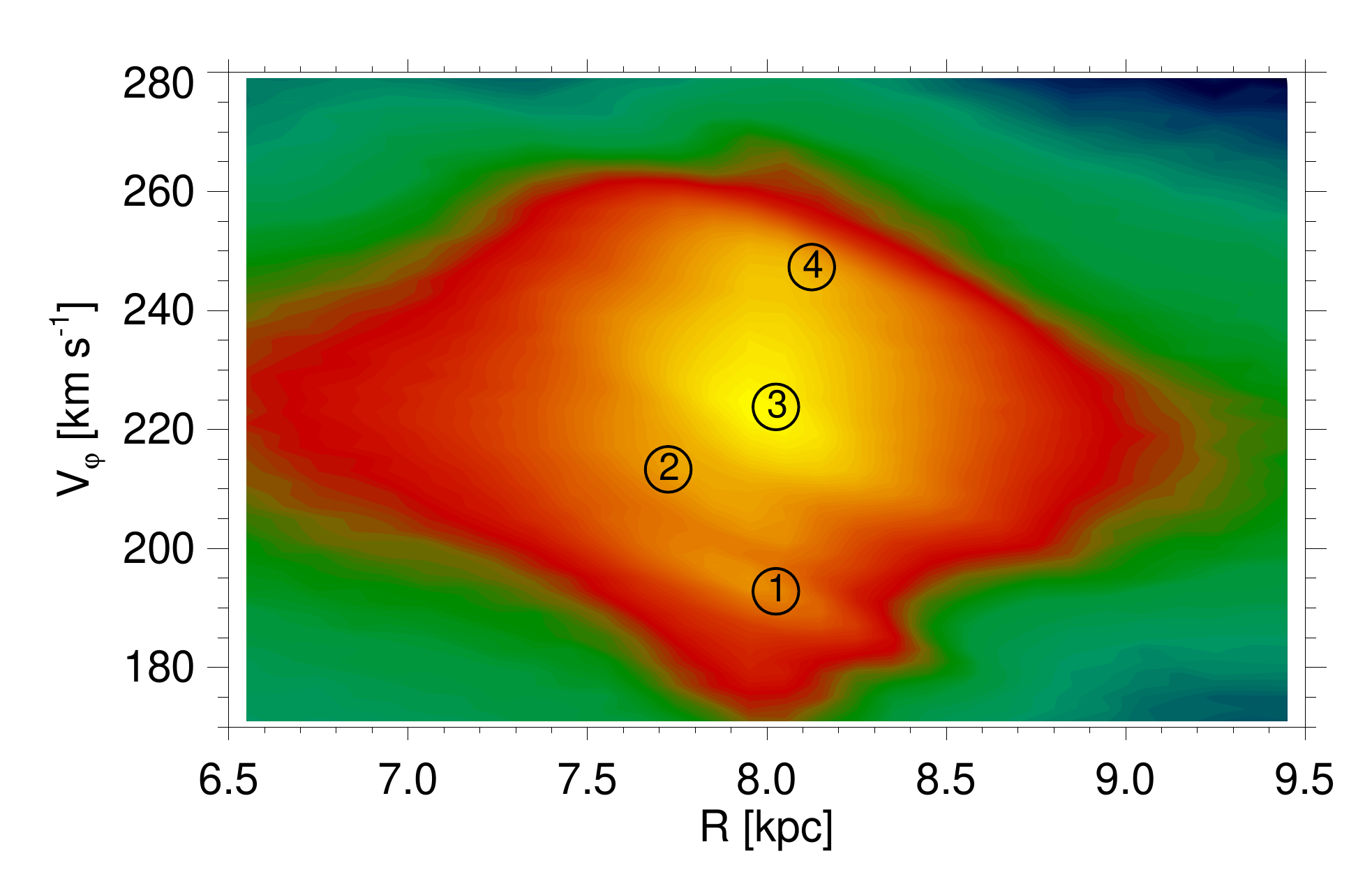}
	\includegraphics[width=1.0\columnwidth]{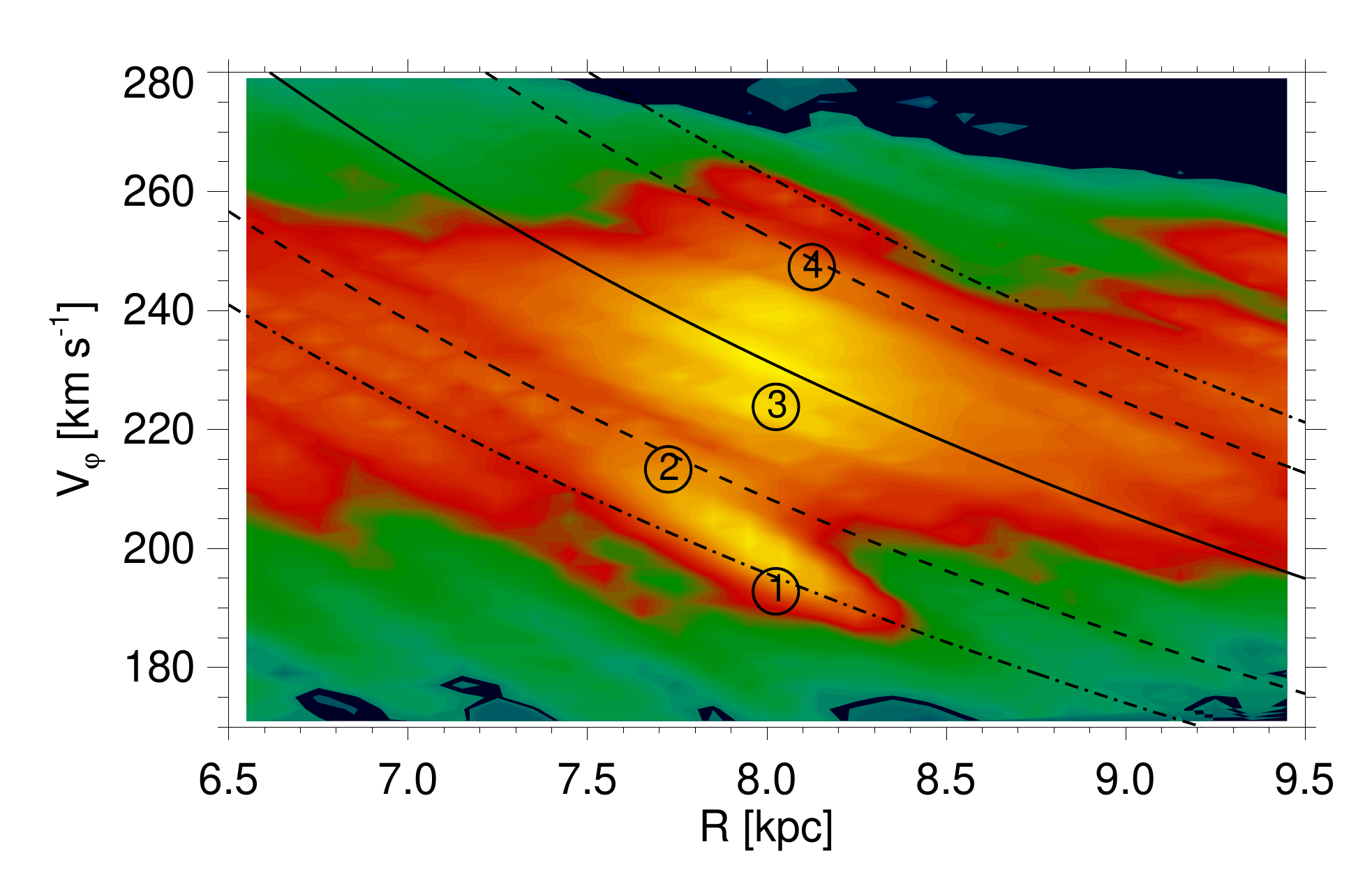}
    \caption{Left: $R$--$V_{\varphi}$ plane of the \textit{Gaia} stars with the logarithm of density mapped by the color scale. The numbers, from 1 to 4, indicate regions of the phase-space associated with some diagonal ridges (see Section~\ref{sec:ridges}). 
    Right: $R$--$V_{\varphi}$ plane resulting from the simulation at $t=1340$\,Myr, with the mapping of the logarithm of density. The numbers, from 1 to 4, and their positions are the same as on the left panel. The black lines indicate families of orbits with constant $L_z$ associated with spiral resonances: solid for the CR, dashed for the 12/1 ILR (lower) and OLR (upper), and dot-dashed for the 8/1 ILR (lower) and OLR (upper).}
    \label{fig:galac_ridges}
\end{figure*}

Here, we focus on the distribution of tangential velocities $V_{\varphi}$ (with respect to the inertial frame) as a function of the radius $R$. 
Figure~\ref{fig:galac_ridges} shows, in the left panel, the distribution of \textit{Gaia} stars on the $R$--$V_{\varphi}$ plane. We selected stars within distances of 4\,kpc from the Sun, at heights from the disk plane of \mbox{$|Z|\leq 500$}\,pc, and with vertical velocities in the range \mbox{$|V_{z}|\leq 20$}\,km\,s$^{-1}$. This selection is intended to work with a sample of stars whose orbits do not depart too much from the Galactic plane, which is suitable for comparison with simulated 2D orbits. The density of stars is shown by a logarithmic color scale in Figure~\ref{fig:galac_ridges}. By numbers, from 1 to 4, we identify some diagonal ridges whose projections on the local $V_R$--$V_{\varphi}$ plane, or equivalently, the local $U$--$V$ plane, connect to density structures associated with moving groups and streams. Number 1 indicates the ridge associated with the Hercules stream; number 2 points to the ridge associated with a weak stream between Hercules and Pleiades-Hyades streams (represented as ``zone III'' in Figure 1 of \citealt{Michtchenko_etal2019}); 
number 3 indicates the ridge associated with the Pleiades-Hyades stream; and number 4 shows the ridge responsible for the Sirius moving group in the local $U$--$V$ plane. 

The right panel of Figure~\ref{fig:galac_ridges} shows the map of the logarithm of density on the $R$--$V_{\varphi}$ plane from the simulation at the time-step \mbox{$t=1340$}\,Myr. The set of numbers, from 1 to 4, is the same as on the left panel. We see a good agreement between the locations of some ridges from the simulated and the observed planes, at least in the local volume space. The black lines represent curves of constant angular momentum $L_{z}$ (\mbox{$V_{\varphi}=L_{z}/R$}); the $L_z$--levels are associated to the nominal positions of the LRs and the CR, whose calculation is described in \cite{Michtchenkoetal2017AA}. The solid curve indicates the CR, and the dashed and dot-dashed curves indicate both 12/1 and 8/1 ILRs (lower) and OLRs (upper), respectively. We can see that the ridge associated with the Hercules stream (indicated by number 1) is formed mainly by orbits trapped around the 8/1 ILR, while the ridge indicated by number 2 is formed by orbits trapped around the 12/1 ILR. The CR is associated to the ridge responsible for the Pleiades, Hyades, and also Coma Berenices groups (indicated by number 3). The ridge that forms the Sirius group (indicated by number 4) is associated to the bulk of overlapping 8/1, 12/1 and higher-order OLRs. These results are in perfect agreement with the findings from Paper I \citep[see also][]{Michtchenko_etal2019}. Nevertheless, a better match of the positions of the diagonal ridges on the observed $R$--$V_{\varphi}$ plane with the ones on the simulated plane can be obtained with slight variations of the locations of the resonances, which can be achieved by adjusting some parameters of the model, in particular, the value of $\Omega_{\mathrm{p}}$ and/or the Galactic parameters $R_0$ and $V_0$.

By comparing the results of our simulations (Figure~\ref{fig:galac_ridges} right panel) with the
structures observed by \citet[][their Figure 6]{Ramos_etal2018}, we see a close correspondence between the main features. For instance, their dashed grey line that passes around $R=8$\,kpc and  $V_{\varphi}=240$\,km\,s$^{-1}$ and follows the densest features is the same level that we identify  with corotation (solid line in Figure~\ref{fig:galac_ridges}). We do not see any strong feature that would require a different explanation than the corotation resonance
of the spiral structure and the myriad of high-order resonances that normally come associated with corotation. 
So, the main structures seen in Figure~\ref{fig:galac_ridges} (left panel) are explained. 
At distances larger than 400\,pc from the Sun, the star counts of the \textit{Gaia} sample drops considerably, and at those regions the association of the ridges with the spiral resonances still has to be investigated, possibly using another technique able to detect low-signal structures like wavelets, as done by \cite{Ramos_etal2018}. However, this is beyond the scope of the present work. 


\section{ Stellar birthplaces}
\label{sec:birthplaces}

In this Section, we investigate the initial positions of the star-particles that populate the SN at the time-step of 1340\,Myr. With this, we can gain some insight on the possible sites of birth of the stars that are currently close to the Sun.

\begin{figure}
	\includegraphics[width=1.0\columnwidth]{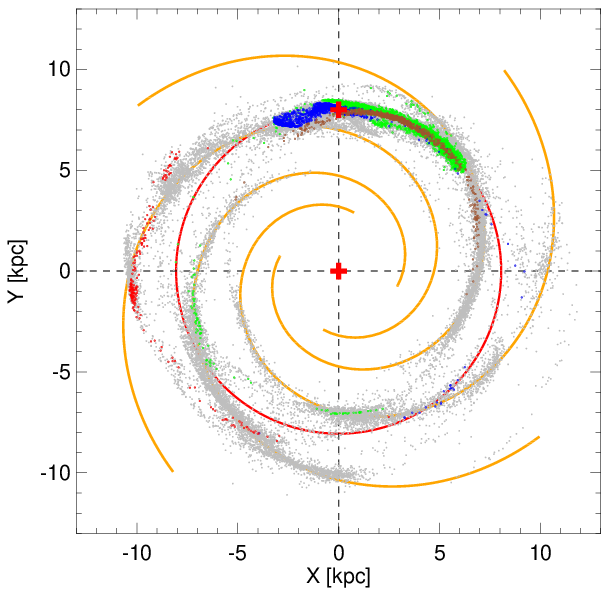}
	\includegraphics[width=1.0\columnwidth]{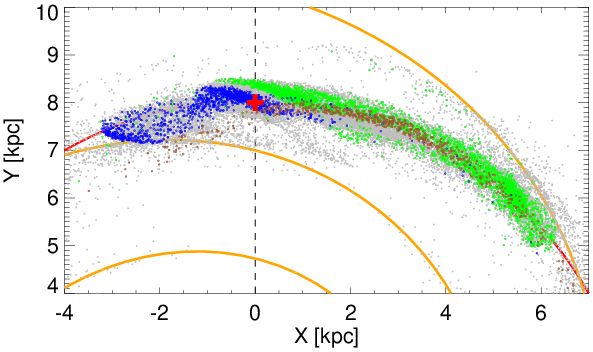}
    \caption{Top: Galactic equatorial $X$--$Y$ plane with initial positions of the star-particles from the simulated $U$--$V$ plane at $t=1340$\,Myr (gray dots). The colored dots correspond to known moving groups and dynamical streams: blue dots for the Coma Berenices group, green dots for the Pleiades-Hyades stream, brown dots for the Hercules stream, and red dots for the Sirius group. The other features on the plane are the same as from Figure~\ref{fig:sim_XY_0}. Bottom: a zoom of the region of the top panel occupied by the Local Arm, for the purpose of better visualization.}
    \label{fig:sim_XY_1}
\end{figure}

Figure~\ref{fig:sim_XY_1} (top panel) shows the $X$--$Y$ plane with the initial positions of all the star-particles that fall within the circle of 150\,pc around the Sun at the time-step \mbox{$t=1340$}\,Myr (gray dots). Besides that, we analyze the initial positions of the members of the moving groups on the simulated $U$--$V$ plane, in particular, the regions associated with the Hercules stream, the Coma Berenices, the Pleiades-Hyades, and the Sirius groups observed on the $U$--$V$ plane from \textit{Gaia}. The colored symbols in Figure~\ref{fig:sim_XY_1} show the initial positions of these star-particles: blue for the Coma Berenices group, green for the region occupied by the Pleiades-Hyades group, brown for a region inside the Hercules stream, and red for the star-particles associated with the Sirius group. It can be noted that the majority of the star-particles that constitute the currently observed Coma Berenices, Hyades, and Pleiades groups in the SN have initial positions along the simulated LA. We can distinguish preferential initial sites of birth; in the bottom panel of Figure~\ref{fig:sim_XY_1} we show a zoom of the region occupied by the LA: the Coma Berenices star-particles are initially mainly along the LA and with Galactic longitudes \mbox{$l>180^{\circ}$}, while the Pleiades-Hyades star-particles are preferentially at longitudes \mbox{$l<180^{\circ}$}. Both of these groups of star-particles have $U$ and $V$ velocities inside the corotation zone in the velocity plane, implying that their orbits have been trapped inside the CR for all the integration timespan of the simulation. According to Paper II, the majority of stars that composes the LA have orbits trapped inside the local corotation zone, in a librating motion between the Sagittarius-Carina and Perseus arms. These pieces of evidence indicate that the stars that form the main component of moving groups on the observed $U$--$V$ plane were born in the LA.

The initial positions of the star-particles selected in the region of Hercules stream are mainly along the LA, but we can also observe a portion of the particles along the inner segment (inside the solar circle) of the Perseus arm. Analyzing the distribution of the initial radial velocities of the star-particles, we obtained that their dispersion in the Hercules group is 2 to 6 times greater than those of the moving groups of the main component. This suggests that, although the star-particles in the Hercules group were placed initially along the LA, their orbits have energies lower than those necessary to evolve inside the zone of influence of the CR, and subsequently are trapped by the nearby high-order ILRs, mainly the 8/1 ILR.

Figure~\ref{fig:sim_XY_1} (top panel) also shows that the initial positions of the star-particles in the Sirius group (red points in the figure) are mainly along an outer branch (outside the solar circle) of the Sagittarius-Carina arm. These star-particles have orbits trapped by the OLRs 8/1, 12/1, 16/1 (Paper I), and visit periodically the SN with positive heliocentric $V$ velocities
during their radial excursions towards the inner Galactic region. 
The above-mentioned outer branch of the Sagittarius-Carina arm, in the third and fourth Galactic quadrants, with Galactic longitudes in the range \mbox{$225^{\circ}\lesssim l \lesssim 280^{\circ}$}, is compatible with Figures 5 and 6 of \cite{Hou_Han2014}, where a number of intense star-forming regions are located, such as H\,{II} regions, giant molecular clouds, and maser sources.


\section{Discussion and Conclusions}
\label{sec:conclusion}

We perform simulations, placing stars in their natural birthplaces, i.e., in the four main spiral arms, along the Local Arm, and in the axisymmetric background disk. By integrating the orbits, allowing the stars to evolve naturally in the 2D Galactic gravitational potential, we recover which birthplaces produce the stars that are presently in the SN. Our simulations reproduce well the large scale structures of the \textit{Gaia} picture of the $U$--$V$ velocity distribution in the SN and, in addition, we provide information on the probable 
origin of the stars, i.e., their initial position and velocity distributions, bringing the long-waited satisfactory understanding of the nature of the moving groups. The velocity distribution of the SN gets its shape due to the peculiarity of the spiral corotation zone, situated close to the Sun, as well as to the high-order LRs that exist near the CR. 

According to the initial conditions and the Galactic model adopted in the present work, the observed velocity structure of the SN is reached in a time-scale of order of 1\,Gyr that puts a dynamical constraint for a long-living nature of the Galactic spiral structure. Thus, our results counterbalance some current arguments in favour of short-lived structures in the SN caused by the ongoing phase mixing of the Galactic disk, expected to have originated from some external event \citep{Antoja_etal2018}. Contrarily, we have shown that the velocity structures observed in the SN, such as the moving groups, arches, streams, ridges, etc., are formed by the action of resonances of the spiral arms on the stellar orbits, capturing stars inside the resonance zones, in a process that lasts, at least, 1\,Gyr for orbits close to the corotation zone.

The simulations also reproduce, with a satisfactory agreement, the diagonal ridges observed in the plot of the $R$--$V_{\varphi}$ plane with \textit{Gaia} data \citep{Antoja_etal2018}. According to our model, the ridges are formed by stellar orbits trapped by the spiral resonances, which form chains that follow curves of constant angular momentum $L_z$ in a dynamical map on the $R$--$V_{\varphi}$ plane of initial conditions \citep{Michtchenkoetal2017AA}. Resonances modify the dynamics in their environment, capturing and trapping stars inside of the stable resonant zones (which enhances the corresponding stellar density), and creating regions of low density close to saddle points and separatrices (Paper I).

Our simulations do not consider a continuous star formation in the disk, but only stars born at the instant \mbox{$t=0$}. In this way, on the simulated $U$--$V$ planes, all the star-particles have the same `age' that is equal to the time-step used for the construction of the plane. This is far from what would be expected and even observed in the distributions of the ages of the stars in the moving groups. \citet[][their Figure 14]{Antoja_etal2008} show such distributions. Apart from some features seen in the distributions that can be explained by the selection of stars for the construction of their sample, like the double main peaks of young and old stars in the groups of Pleiades, Hyades, Coma Berenices and Sirius, what is evident is that very young stars come from the main component of moving groups, not from the Hercules group. According to our results, the majority of these young stars were born in the LA, with positions and velocities that put them inside the stable zone of the CR. Besides that, the peak at \mbox{$\sim 2$}\,Gyr seen in the age distributions of the above-mentioned moving groups, as well as in the Hercules group, is compatible with our results for the time of appearance and stabilization of the structures on the $U$--$V$ plane. The older stars in the age distribution given in \citet[][their Figure 14]{Antoja_etal2008} provide a lower limit for the age of the moving groups, and consequently, for the formation of the spiral structure, that would be about 3--4\,Gyr.

One of our interesting results is that most stars of the main region of the CR were born in this same region, in which they are trapped. A possibly related fact comes from studies of the age-metallicity distribution of the stars of the SN \citep{Nordstrom_etal2004,Holmberg_etal2007}, which show that there has been a relatively fast increase in metallicity during the last 3\,Gyr. The high metallicity of the corotation region is also seen by \cite{Spina_etal2017}. The coincidence of star trapping and high metallicity in a same Galactic region lets us to speculate about the nature of the connection.

Regarding the LA, the story of the stellar population in there suggests 
that the corotation region could have been the site of a chemical evolution that is not far from a closed-box system. In such a model, the metallicity increases rapidly with time, because the new stars are formed from the material of the interstellar medium which is continuously enriched in metals, due to the return of mass from the massive stars that reach the end of their life. The high metallicity of the corotation region could explain the existence of a metallicity step at the outer frontier of the corotation zone, as well as the azimuthal gradient \citep{Lepine_etal2011}. These are possibly local phenomena associated with the resonant zone, and not an axisymmetric radial gradient of the whole Galaxy. The hypothesis discussed here suggests, therefore, that the age of the corotation zone could be about 3\,Gyr.


\section*{Acknowledgements}

We acknowledge the anonymous referee for the detailed review and for the helpful suggestions, which allowed us to improve the manuscript. This work was supported by the Brazilian agencies Conselho Nacional de Desenvolvimento Cient\'ifico e Tecnol\'ogico (CNPq), the Coordena\c{c}\~ao de Aperfei\c{c}oamento de Pessoal de N\'ivel Superior - Brasil (CAPES) - Finance Code 001, and the S\~ao Paulo Research Foundation (FAPESP) (grant 2016/13750-6). R.S.S.V. acknowledges CAPES for financial support. A.P.V. acknowledges FAPESP for the postdoctoral fellowship No. 2017/15893-1 and the DGAPA-PAPIIT grant IG100319. This work has made use of data from the European Space Agency (ESA) mission
{\it Gaia} (\url{https://www.cosmos.esa.int/gaia}), processed by the {\it Gaia}
Data Processing and Analysis Consortium (DPAC,
\url{https://www.cosmos.esa.int/web/gaia/dpac/consortium}). Funding for the DPAC
has been provided by national institutions, in particular the institutions
participating in the {\it Gaia} Multilateral Agreement.




\end{document}